\documentclass[10pt]{iopart}
 \pdfoutput=1
\usepackage{iopams}
\usepackage{epsfig,cite}
\newcommand{\be}{\begin{equation}}
\newcommand{\ee}{\end{equation}}
\newcommand{\bd}{\begin{displaymath}}
\newcommand{\ed}{\end{displaymath}}
\newcommand{\BE}{\begin{eqnarray}}
\newcommand{\EE}{\end{eqnarray}}

\begin{document}

\title{Optimizing evacuation flow in a two-channel exclusion process}

\author{Tobias Galla}

\address{
Theoretical Physics, School of Physics and Astronomy, The University of Manchester, Manchester M13 9PL, United Kingdom
}

\begin{abstract}
We use a basic setup of two coupled exclusion processes to model a stylised situation in evacuation dynamics, in which evacuees have to choose between two escape routes. The coupling between the two processes occurs through one common point at which particles are injected, the process can be controlled by directing incoming individuals into either of the two escape routes. Based on a mean-field approach we determine the phase behaviour of the model, and analytically compute optimal control strategies, maximising the total current through the system. Results are confirmed by numerical simulations. We also show that dynamic intervention, exploiting fluctuations about the mean-field stationary state, can lead to a further increase in total current.

\end{abstract}


\ead{\tt Tobias.Galla@manchester.ac.uk}


\section{Introduction}
There are in general several lines of approaching the individual-based modelling of complex human interaction. Social scientists tend to use detailed models aiming to realistically map out as many aspects of the real-world dynamics as possible. This school of thought favours realism over simplicity\footnote{Some social scientists go so far as to say `Simplicity is no reason, difficulty is no excuse.' \cite{bruce}.}, and is often referred to as `agent-based modelling', see for example \cite{bonabeau} and references therein. It typically leads to complex simulations, and models with a large number of parameters. The approach taken by physicists on the other hand does not aim to provide models which are realistic in all detail, but instead tries to capture only the features absolutely essential for the underlying dynamics. This favours simplicity over realism, and generally tends to produce stylised models with only a small number of parameters. The advantage of the detailed modelling approach is that it tends to capture the real-world dynamics to a higher accuracy, making it potentially easier to make quantitative predictions. The number of parameters in such models, on the other hand, can make it difficult to distinguish between cause and effect and to unravel the precise mechanisms at work. The physicists' models at the other end of the spectrum are often so simple that it is difficult to make quantitative statements about specific applications. Their simplicity, however, frequently makes it possible to obtain analytical solutions (either exact or in good approximation), hence the outcome of the model dynamics can often be calculated without the need for extensive simulations. Furthermore the relatively small number of parameters allows for a more exhaustive analysis. Simple models of human interaction are at the core of what is now called complexity science, owing mostly to the realization that simple dynamics on the micro-level can result in complex behaviour on the macro-level.

\begin{figure}[t]
\vspace{0em}
\begin{center}
\includegraphics[width=1.\textwidth]{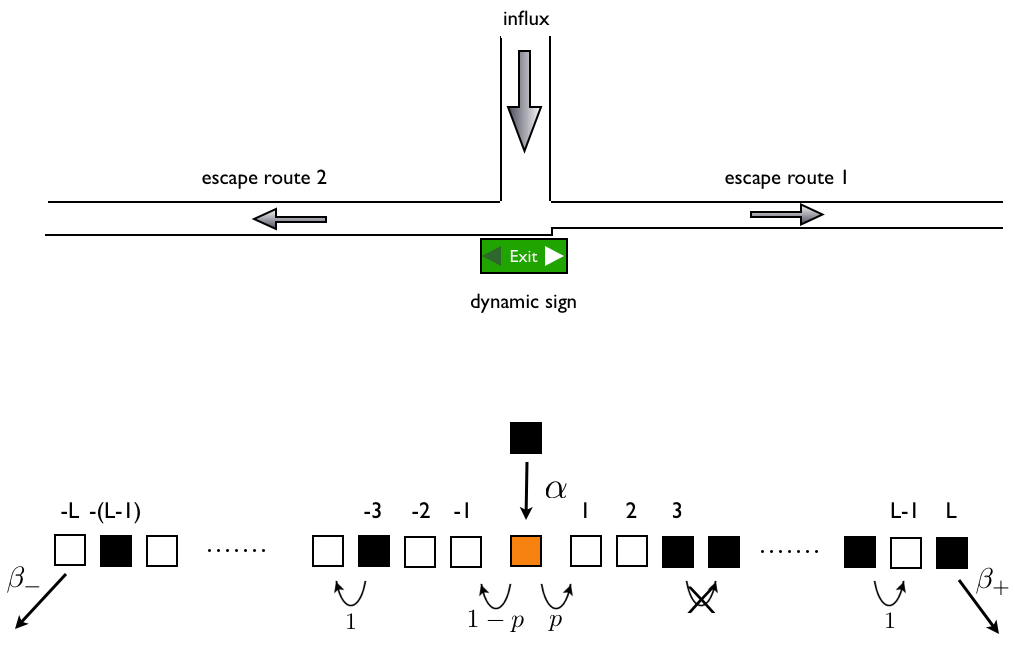}
\end{center}
\caption{\label{fig:fig1} Illustration of the model: Upper panel shows an evacuation scenario with a T-junction, with an influx of evacuees from the top, who then have to decide whether to move into the left-hand or right-hand corridor. A dynamic sign, directing individuals to the left or right, is placed at the junction, and may be adapted.  Lower panel shows an abstraction of this situation, based on two asymmetric exclusion processes, coupled through a central site. Particles injected into the central site decide to move into the two branches with probabilities $p$ and $1-p$ respectively. The control parameter $p$ can be adapted to optimise the flux given model parameters (injection rate $\alpha$, ejection rates at the end of the branches, $\beta_+$ and $\beta_-$).}
\end{figure}

In this paper we take the physicists' approach to modelling complex systems, and investigate a simple model of evacuation dynamics with dynamic signage. More precisely we study the stylized situation depicted in the upper panel of Fig. \ref{fig:fig1}. Evacuees are assumed to enter a T-junction via one corridor, and then have to choose between one of two exit routes, each one with different properties. A dynamic sign is placed at the junction, and allows emergency managers to direct incoming individuals to either of the two exit routes. We do not assume that controllers can force individuals to move to the right or to the left, instead we consider a more general model in which the  control parameter is the probability $p$ of individuals choosing escape route $1$, versus the probability $1-p$ of choosing escape route $2$\footnote{Consistent with the stylised modelling approach taken in our paper we do not specify how exactly a controller may vary the probability for particles moving to the left/right {\em quantitatively} in practice. Instead what we assume is that incoming particles follow the direction indicated by a dynamic sign probabilistically, i.e. some may choose the indicated escape route, others may not.}.
The actual model dynamics is defined by the rules by which evacuees are propagated in space. We here consider one of the most stylised models available, and focus on the so-called totally asymmetric exclusion process (TASEP) \cite{macdo1,macdo2, derrida, derrida2}. This process defines a cellular automaton with one basic update rule: if an agent is picked for update they move to the cell in front of them if and only if this cell is vacant. Particles are injected with rate $\alpha$ at the T-junction, and removed from the system with rates $\beta_+$ and $\beta_-$ at either end of the two escape routes. The precise mathematics of the update algorithm will be specified in the next section.
While this model is undoubtedly a very minimalist model of pedestrian motion, a significant body of literature is available on TASEP models. In particular full solutions of the different possible phases in single exclusion processes have been worked out \cite{derrida}, and are hence ready for use in models of coupled processes. Asymmetric exclusion processes have not only been used to model pedestrian dynamics \cite{wolki,helbing}, they are also related to models of vehicular traffic \cite{nagel,schad,schad2,helbing}, and they are popular in the modelling of biological transport processes as well \cite{macdo1,macdo2, shaw, dong,frey}. From the point of view of statistical physics they constitute an interesting class of driven lattice gases off equilibrium \cite{evans}. Exclusion processes have been studied on the mean-field level, but an increasing interest is also on their stochastic dynamics and large fluctuation behaviour \cite{zia1,zia2}. ASEP and TASEP models with disorder have been studied \cite{harris,santen}, and symmetry-breaking has been seen to occur in systems of coupled exclusion processes \cite{kolo1}. In this paper we investigate the phase behaviour of two coupled TASEPs with one central control parameter (the variable directing the incoming flux into either of the two branches of the system). In particular we ask what the optimal strategy is to maximise the total current through the system, given the injection rate of particles, and the ejection rates at the end of the two branches of the system. We show that based on a mean-field approach the total current can be increased significantly over a naive approach based on the ratio of ejection rates. We also show in simulations that dynamic control can lead to a further increase in performance.

The remainder of the paper is structured as follows:  Sec. \ref{sec:model} contains the specifics the model dynamics. In Sec. \ref{sec:mf} we work out the mean-field theory of the model, study the different possible phases of the coupled system, and derive explicit equations for the total current through the system. Sec. \ref{sec:control} focuses on the optimal control of the system, assuming adiabatically slow intervention. In Sec. \ref{sec:dyncontrol} we finally address dynamic intervention strategies, in which control parameters can be varied in time depending on the dynamical state of the system. Sec. \ref{sec:summary} contains our conclusions and an outlook to future lines of research.

\section{Model}\label{sec:model}
An abstraction of the evacuation scenario depicted in the upper panel of Fig. \ref{fig:fig1} can be obtained by considering two TASEPs, coupled via one central site, as shown in the lower panel of Fig \ref{fig:fig1}.  More precisely, there are $N=2L+1$ sites in our model, each of which can be occupied or vacant. We label sites by $i=-L,-(L-1),\dots,-1,0,1,\dots,L-1,L$. The central site is labelled $i=0$. The right-hand branch consists of the sites $i=1,\dots,L$, the left-hand escape route of the sites $i=-1,\dots,-L$. Occupation variables $n_i(t)\in\{0,1\}$ indicate whether or not a given cell is occupied ($n_i(t)=1$) or vacant ($n_i(t)=0$) at time $t$. Particles are injected at the central site at a rate $\alpha$, i.e. if site $i=0$ is found to be empty it is filled with probability $\alpha$ when the central site is chosen for update (see below for further details). Removal of particles occurs at the end of the two branches, with rates $\beta_+$ and $\beta_-$ respectively, i.e. if site $\pm L$ is found to be occupied the corresponding particle is removed with probability $\beta_\pm$. The injection rate $\alpha$ and the ejection rates $\beta_\pm$ constitute the main parameters characterizing the physical layout of the structure. They take values in the interval $[0,1]$. A fourth model parameter is given by $p\in[0,1]$, and models a control strategy, which can be modified adapting the signage at the junction: incoming particles, placed at site $i=0$, `decide' to travel into the right-hand branch with probability $p$, and into the left-hand branch with probability $1-p$. It is here important to stress that we consider a rather simple model, in which the decision whether to travel to the right or the left is made immediately upon arrival at the central site. I.e. agents cannot stay put at the central site for a while and then decide which exit route to take depending on observations they may make of the state of the system. Our agents are therefore zero-intelligence individuals, all that can be controlled externally is the probability $p$ with which they choose the right-hand branch. The system is initialized in an empty state, $n_i(t=0)=0$ for all $i$. The dynamics thereafter can then be summarized as follows:
\begin{enumerate}
\item[1.]  Increment time by $\Delta t=1/(2L+1)$. Pick one site $i\in\{-L,\dots,L\}$ at random.
\item[2.] If $i=0$ and $n_i=0$ then 
\begin{itemize} 
\item[(a)] with probability $\alpha$ the central site is filled ($n_0=1$) and a direction of motion $d\in\{+1,-1\}$ is chosen: $d=+1$ with probability $p$ and $d=-1$ with probability $1-p$ respectively.  Any previous choice of $d$ is erased as the corresponding particle is no longer at the central site. Go to 1.
\item[(b)] with probability $1-\alpha$ the value of $n_i$ remains zero. Go to 1.
\end{itemize}
\item[3.] If $i=0$ and $n_i=1$ then check whether site $d$ is occupied (with $d\in\{-1,+1\}$ the direction of motion chosen when site $0$ was filled). If site $d$ is vacant, then set $n_0=0$ and $n_d=1$. Otherwise do nothing. Go to 1. 
\item[4.] If $i$ is in the bulk, i.e. $i\neq 0$ and $|i|\neq L$, 
\begin{itemize}
\item[(a)] if $i>0$, $n_i=1$ and $n_{i+1}=0$ set $n_{i+1}=1, n_i=0$ (particle hops to the right). Goto 1.
\item[(b)] if $i<0$, $n_i=1$ and $n_{i-1}=0$ set $n_{i-1}=1, n_i=0$ (particle hops to the left) Goto 1.
\item[(c)] in all other cases do nothing. Go to 1.
\end{itemize}
\item[4.] If $i=L$  and $n_L=1$ then with probability $\beta_+$ set $n_L=0$, with probability $1-\beta_+$ the site remains occupied. Go to 1.
\item[5.] If $i=-L$ and $n_{-L}=1$ then with probability $\beta_-$ set $n_{-L}=0$, with probability $1-\beta_-$ the site remains occupied. Go to 1.
\item[6.] If $i=\pm L$, and $n_i=0$ do nothing. Go to 1.
\end{enumerate}
We have here chosen the ejection rates $\beta_\pm$ to model heterogeneity between the two exit routes, i.e. their properties will differ whenever $\beta_+\neq \beta_-$. The ejection rates can here be related to filling factors of reservoirs at the end of each branch, similar to constrained exclusion processes discussed e.g. in \cite{harris,zia1,zia2}. In an evacuation scenario these reservoirs may be thought of as shelters or similar. The choice of using $\beta_\pm$ to model the properties of the two branches of the evacuation system was here made mainly for convenience. Other choices are possible, for example one could assume that particles move forward in one branch with unit probability if the site ahead is vacant, but with a smaller probability in the other branch. Alternatively the lengths of the two branches, as measured by the number of cells in each branch, could be taken to differ (this latter choice would however not be captured by a mean-field approach).
\section{Mean-field analysis}\label{sec:mf}
\subsection{Phases of the standard TASEP model}\label{sec:stasep}
Within mean-field theory the standard TASEP model with an injection rate $\alpha$ and an ejection rate $\beta$ is known to exhibit three different phases \cite{derrida}:
\begin{enumerate}
\item Low-density phase ($\alpha\leq1/2, \beta>\alpha$): the first cell along the chain is occupied with probability $\rho_0=\alpha$, the current is given by $j=\alpha(1-\alpha)$.
\item High-density phase ($\beta\leq1/2, \alpha>\beta$): the first cell is occupied with probability $\rho_0=1-\frac{\beta(1-\beta)}{\alpha}$, the current is given by $j=\beta(1-\beta)$.
\item Maximum-current phase ($\alpha\geq1/2, \beta\geq1/2$): the first cell is occupied with probability $\rho_0=1-\frac{1}{4\alpha}$, the current is $j=1/4$.
\end{enumerate}
There is also a forth `coexistence phase' \cite{derrida}', which is realised when $\alpha=\beta\leq 1/2$. Since this phase is only observed for rather special choices of parameters (forming a set of measure zero in parameter space), we will focus our attention on the above three phases, all realized for a large range of parameters. 
\subsection{Relations governing the coupled model}
It is convenient to view the present setup as a model coupling two TASEPs. The coupling is through the central site $i=0$, which can only be occupied by one particle at any time. If this site is occupied a direction of motion is specified for the particle occupying this site (see the above algorithm). Thus, until the central particle has moved and the central site has become vacant again, no particles can be injected into either branch of system. Our analysis focuses on the stationary state, i.e. we assume constant currents $j_+$ and $j_-$ through the two branches of the system. Let us denote the probability for site $i\in\{-L,\dots,L\}$ to be occupied in the stationary state by $\rho_i$. As discussed above the central site can either be occupied by a particle intending to travel to the right, or by a particle intending to travel to the left. We assume that this occurs with probabilities $\rho_0^+$ and $\rho_0^-$ respectively, so that we have $\rho_0=\rho_0^++\rho_0^-$. Within the mean-field description the total influx into the system is given by $j=\alpha(1-\rho_0)$, where the flux is defined as the number of particles entering the system (equivalently travelling through the system) per unit time.  We therefore have
\be\label{eq:b1}
\alpha(1-\rho_0)=j_++j_-.
\ee
The second conservation law we will be using is given by
\be\label{eq:b2}
p\alpha(1-\rho_0)=j_+,
\ee
indicating that a fraction $p$ of the total influx enters the positive branch of the system, and hence equals the current in that branch\footnote{Equivalently we could use the corresponding relation, $(1-p)\alpha(1-\rho_0)=j_-$, for the negative branch.}. Eq. (\ref{eq:b1}) and (\ref{eq:b2}) will be the starting point of our analysis. We will proceed by assuming that particles enter the positive branch with an {\em effective} injection rate $\alpha_+$, and the negative branch with an effective injection rate $\alpha_-$ (i.e. if chosen for update and found not to contain a particle with motion in direction $\pm 1$ then the central site will be filled with such a particle with probability $\alpha_\pm$). It is here important to stress that these rates are not model parameters, but that instead they are self-consistently determined by the dynamics of the system, once the control parameters $\alpha,\beta_+,\beta_-$ and $p$ have been chosen.

Depending on the values of the model parameters $\beta_\pm$, and the values of the effective injection rates $\alpha_\pm$, dictated by the above conservation and self-consistently relations, each branch can be in one of the three phases of a single TASEP, allowing in principle for nine possible phases of the combined system. We will consider these separately in the following, deriving conditions for when they can be realised. Phases such as HL (positive branch in HD phase, negative branch in LD phase) and the reverse situation, LH, are related by a simple interchange of indices, so that only one combination will be considered in such cases.

\subsection{High-high phase (HH)}
If both branches of the system are in their HD phases, then we have 
\be
\rho_0^\pm=1-\frac{\beta_\pm(1-\beta_\pm)}{\alpha_\pm}, ~~ j_\pm = \beta_\pm(1-\beta_\pm).
\ee
Using this in Eqs. (\ref{eq:b1}) and (\ref{eq:b2}) one concludes
\be
p=\frac{\beta_+(1-\beta_+)}{\beta_+(1-\beta_+)+\beta_-(1-\beta_-)}.
\ee
Thus the HH phase can only occur in a zero-measure set of parameter space.
\subsection{Low-High phase (LH)}
Let us assume that the positive branch is in the LD phase, and the negative branch in the HD phase. We then have
\be\label{eq:lhrho}
\rho_0^+=\alpha_+, ~~~~\rho_0^-=1-\frac{\beta_-(1-\beta_-)}{\alpha_-}
\ee
as well as
\be\label{eq:lhj}
j_+=\alpha_+(1-\alpha_+), ~~~~ j_-=\beta_-(1-\beta_-).
\ee
Inserting this into Eqs. (\ref{eq:b1},\ref{eq:b2}) one finds
\BE
\alpha(1-\rho_0)=\alpha_+(1-\alpha_+)+\beta_-(1-\beta_-),\\
p\alpha(1-\rho_0)=\alpha_+(1-\alpha_+)
\EE
From this one concludes
i.e.
\be\label{eq:rho0hl}
\rho_0=1-\frac{\beta_-(1-\beta_-)}{\alpha(1-p)},
\ee
i.e. we have found an explicit expression for $\rho_0$ in terms of the model parameters. This can then be used to solve
\be
\alpha(1-\rho_0)=\alpha_+-\alpha_+^2+\beta_-(1-\beta_-)
\ee
for $\alpha_+$. Specifically one finds the physical solution
\be\label{eq:alphaphl}
\alpha_+=\frac{1}{2}-\sqrt{\frac{1}{4}+\beta_-(1-\beta_-)-\alpha(1-\rho_0)},
\ee
which in turn determines the current $j_+=\alpha_+(1-\alpha_+)$, and the total current $j=\alpha_+(1-\alpha_+)+\beta_-(1-\beta_-)$. Finally we use the condition
\be
\alpha_+ + 1-\frac{\beta_-(1-\beta_-)}{\alpha_-}=\rho_0
\ee
and the result of Eq, (\ref{eq:rho0hl}) to obtain
\be\label{eq:alphamhl}
\alpha_-=\left[\frac{\alpha_+}{\beta_-(1-\beta_-)}+\frac{1}{\alpha(1-p)}\right]^{-1},
\ee
which can be evaluated explicitly using Eq. (\ref{eq:alphaphl}).
The case in which the negative branch is in the LD phase, and the positive branch in the HD phase can be dealt with analogously.
\subsection{High - Maximum Current phase (HM)}
In this phase we have 
\be
\rho_0^+=1-\frac{\beta_+(1-\beta_+)}{\alpha_+}, ~~~ \rho_0^-=1-\frac{1}{4\alpha_-},
\ee
and
\be
j_+=\beta_+(1-\beta_+), ~~~~ j_-=\frac{1}{4}.
\ee
Using Eqs. (\ref{eq:b1},\ref{eq:b2}) this implies
\be
p=\frac{\beta_+(1-\beta_+)}{\beta_+(1-\beta_+)+\frac{1}{4}},
\ee
which means that this phase cannot be realised in any extended part of parameter space.

\subsection {Low-Low Phase (LL)}
If we assume that both branches are in their LD phases, one has
\BE
\rho_0^+=\alpha_+, ~~~~~ \rho_0^-=\alpha_-,
\EE
as well as
\BE
j_+=\alpha_+(1-\alpha_+), ~~~~~ j_-=\alpha_-(1-\alpha_-).
\EE
Inserting this into Eqs. (\ref{eq:b1},\ref{eq:b2}) one finds
\BE
\alpha(1-\alpha_+-\alpha_-)&=&\alpha_+(1-\alpha_+)+\alpha_-(1-\alpha_-), \\
p\alpha(1-\alpha_+-\alpha_-)&=&\alpha_+(1-\alpha_+)\label{eq:fump}
\EE
Taking the difference of these two equations one gets
\be\label{eq:alphap}
\alpha_+=1-\alpha_-\left[1+\frac{1-\alpha_-}{(1-p)\alpha}\right].
\ee
Inserting this in (\ref{eq:fump}) one has
\be\label{eq:all}
\alpha_-=\alpha_-^2+\frac{1-p}{p}\left[1-\alpha_-\left(1+\frac{1-\alpha_-}{(1-p)\alpha}\right)\right]\alpha_-\left(1+\frac{1-\alpha_-}{(1-p)\alpha}\right),
\ee
which can be solved numerically for $\alpha_-$. The effective injection rate into the positive branch can then be calculated from Eq. (\ref{eq:alphap}), and with both effective injection rates known, we can then compute the total current
\be
j=\alpha_+(1-\alpha_+)+\alpha_-(1-\alpha_-)
\ee
as a function of the model parameters.
\subsection{Maximum current - Low density phase (ML) }
Let us now assume that the positive branch is in the MC phase, and the negative branch in the LD phase. The reverse case can be treated analogously. One then has
\be
\rho_0^+=1-\frac{1}{4\alpha_+}, ~~~~ \rho_0^-=\alpha_-, \label{eq:r2}
\ee
and
\be
j_+=\frac{1}{4}, ~~~~ j_-=\alpha_-(1-\alpha_-).
\ee
Taking into account that the stationary flux into the positive (MC) branch is given by $j_+=p\alpha(1-\rho_0)$ one has
\be
p\alpha(1-\rho_0)=\frac{1}{4},
\ee
i.e.
\be\label{eq:rho0ml}
\rho_0=1-\frac{1}{4p\alpha}.
\ee
From the condition on the total current one finds
\BE
\alpha(1-\rho_0)=\frac{1}{4}+\alpha_-(1-\alpha_-),
\EE
from which we have, using Eq. (\ref{eq:rho0ml}),
\be\label{eq:alpha1ml}
\alpha_-=\frac{1}{2}-\sqrt{\frac{1}{2}-\frac{1}{4p}},
\ee
where again we have selected the physically realised solution. As a final step we use the condition $\rho_0=\rho_0^++\rho_0^-$, and Eqs. (\ref{eq:r2},\ref{eq:rho0ml}) to find
\be
\alpha_+=\left[4\alpha_-+\frac{1}{p\alpha}\right]^{-1},
\ee
where $\alpha_-$ is known from Eq. (\ref{eq:alpha1ml}).
\subsection{MC-MC phase (MM)}
Let us now assume that both branches are in the MC phase. Then one has $j_+=j_-=1/4$, i.e. $j=1/2$. Eq. (\ref{eq:b1})  then dictates $\alpha(1-\rho_0)=1/2$, i.e.
\be\label{eq:rho0mm}
\rho_0=1-\frac{1}{2\alpha}.
\ee
Given that $j_+=1/4$ we conclude from Eq. (\ref{eq:b2}) that $p\alpha(1-\rho_0)=1/4$, so that we must necessarily have $p=1/2$ in order to this phase to be possible. If both branches are in the MC phase we also have 
\be\label{eq:rho0mm2}
\rho_0=\rho_0^++\rho_0^-=2-\frac{1}{4\alpha_+}-\frac{1}{4\alpha_-}.
\ee
From this we conclude
\be
-\frac{1}{2\alpha}=1-\frac{1}{4\alpha_+}-\frac{1}{4\alpha_-},
\ee
which can never be fulfilled as the LHS is negative, and the RHS non-negative for $\alpha_\pm\geq 1/2$ (as required for the two branches to be in the maximum-current state). Therefore there can be no MM phase.
\begin{figure}[t]
\vspace{0em}
\begin{center}
\includegraphics[width=.9\textwidth]{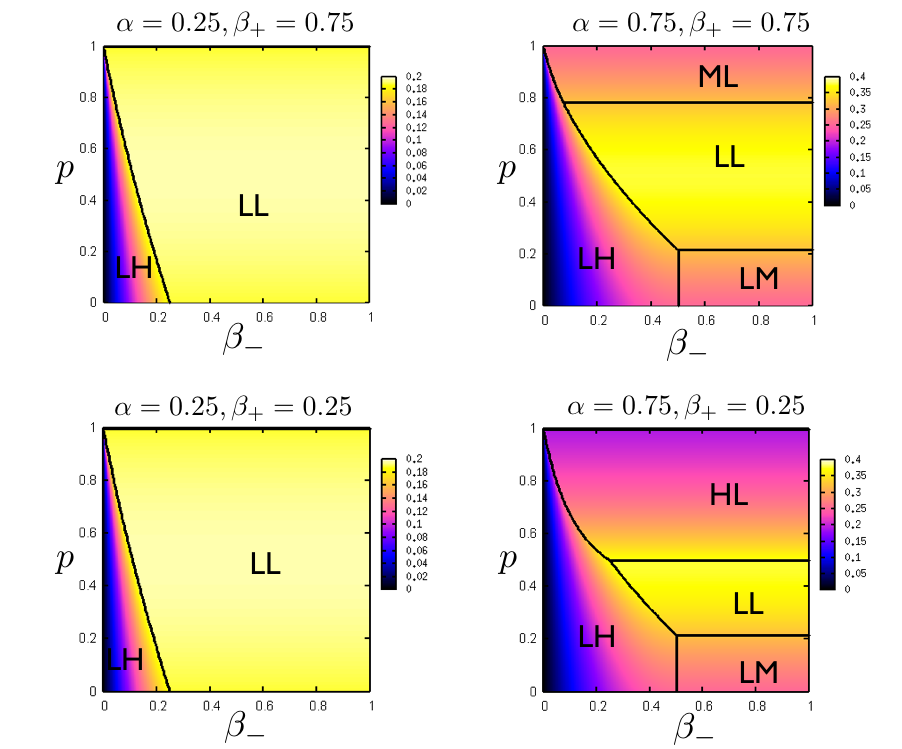}
\end{center}
\caption{\label{fig:pg1} (Colour online) Phase diagram of the model in the $(\beta_-,p)$-plane for different choices of the model parameters $\alpha$ and $\beta_-$. The black lines indicate the phase boundaries, colour in the background represents the magnitude of the total current.}
\end{figure}
\subsection{Phase diagrams}
We have now obtained analytical expressions for the effective injection rates $\alpha_\pm$ and the currents in each of the branches as a function of the model parameters $\alpha,\beta_\pm$ and $p$. These expressions are fully explicit for the LH and LM phases, and implicit for the LL phase. The implicit expression for the LL phase, Eq. (\ref{eq:all}), is of fourth order, and so it can in principle be solved analytically, we have here instead resorted to a numerical solution by iteration. The expressions obtained for $\alpha_\pm$ in each phase can be used to determine range of the individual phases, applying the conditions given in Sec. \ref{sec:stasep}. For example the negative branch can be in the L phase if and only if $\alpha_-\leq1/2, \beta_->1/2$, etc. We have not attempted to determine the phase boundaries analytically. Instead for each point in parameter space we have computed the values of $\alpha_\pm$ one would obtain assuming the combined system is in the LL, LH, HL, LM and ML phase, respectively, and have then checked for consistency with the conditions on $\alpha_\pm,\beta_\pm$ which apply to each of these phases. In all cases we have tested we find that the expressions and conditions of only one phase are valid, hence identifying the physically realised phase diagram uniquely\footnote{Numerical effects can lead to ambiguities at or near the phase lines. Whether or not this occurs can depend on the details of the iterative method used to solve the equations describing the LL phase.}. Examples are shown in Figs. \ref{fig:pg1} and \ref{fig:pg2}. As seen in the figures the model displays quite complex phase behaviour. We here only show a selection of phase diagrams, other topological arrangements of the phases in parameter space may be possible. It is interesting to note that the phase diagrams shown in the left-hand column of Fig. \ref{fig:pg1} appear identical. This is indeed confirmed inspecting the analytical solutions. The expressions for $\alpha_\pm$ in the LH phase (see Eqs. (\ref{eq:alphaphl},\ref{eq:alphamhl})) reveal that $\alpha_\pm$ do not depend on the choice of $\beta_+$, and the same statement holds in the LL phase, see Eqs. (\ref{eq:alphap},\ref{eq:all}). 

It is also interesting to observe the magnitude of the total current as a function of the model parameters. The total flux through the system is indicated as a background colour map in Figs. \ref{fig:pg1} and \ref{fig:pg2}, with lighter colours indicating a higher current than darker colours. The maximum current in each branch is given by $j_\pm=1/4$, see Sec. \ref{sec:stasep}, so the total current can never exceed the value $j=1/2$. As seen in the figures the control parameter $p$ has relatively little influence on the total current if the injection rate $\alpha$ is small (see left column of Fig. \ref{fig:pg1}, and left panel of Fig. \ref{fig:pg2}). For the case of more congested systems, with a high injection rate $\alpha$, and for which at least one branch has a relatively low ejection rate, the question of whether incoming particles are directed to the left or right (as parametrized by $p$) becomes important. In such cases there is generally only a rather small band of values for $p$ yielding an optimal or close-to-optimal total flux (see lower right hand panel of Fig. \ref{fig:pg1}, and right hand panel of Fig. \ref{fig:pg2}).
\section{Test against simulations and optimal control policy}\label{sec:control}
 \begin{figure}[t]
\vspace{0em}
\begin{center}
\includegraphics[width=.85\textwidth]{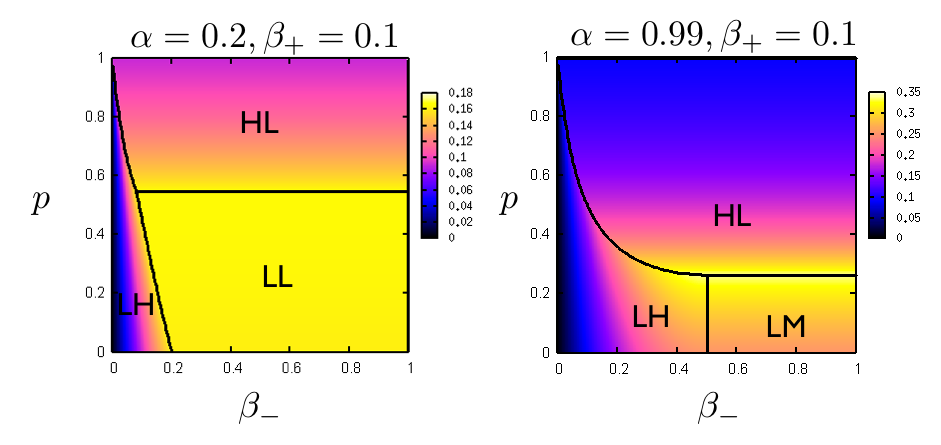}
\end{center}
\caption{\label{fig:pg2}(Colour online) Phase diagram of the model in the $(\beta_-,p)$-plane for different choices of the model parameters $\alpha$ and $\beta_+$. The black lines indicate the phase boundaries, colour in the background represents the magnitude of the total current. }
\end{figure}
\begin{figure}[t]
\vspace{0em}
\begin{center}
\includegraphics[width=.48\textwidth]{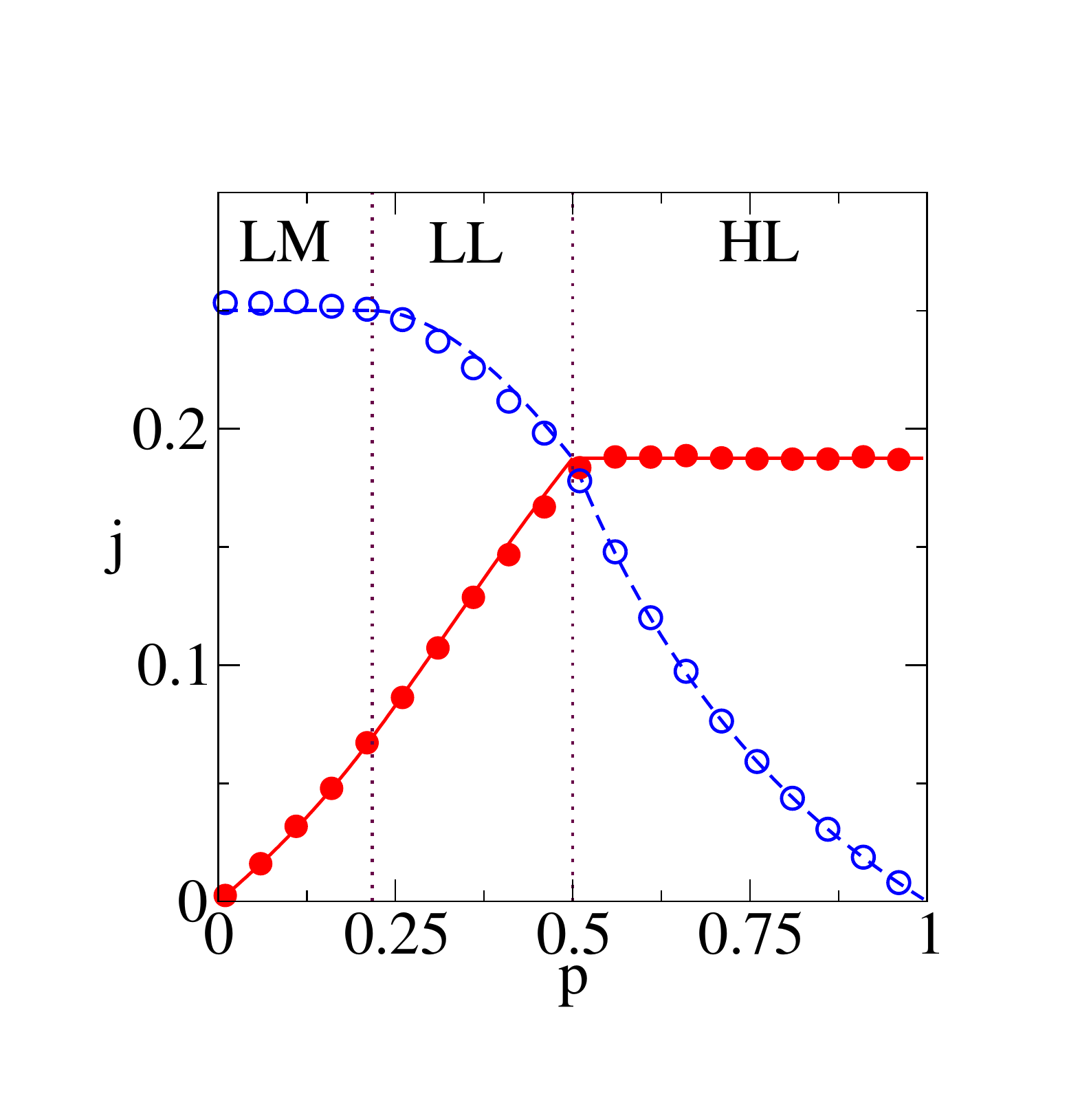} ~~~\includegraphics[width=.48\textwidth]{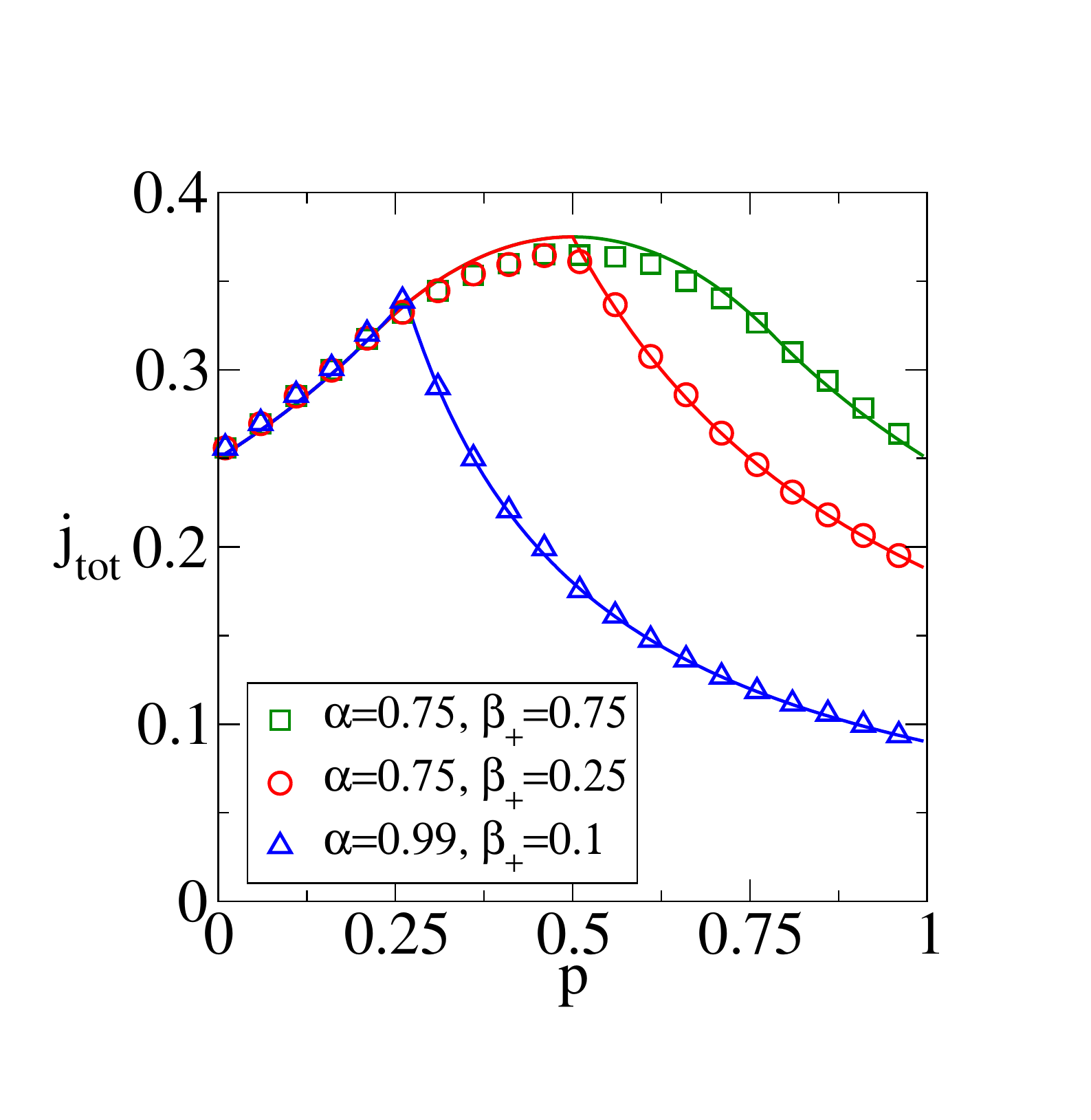} 
\end{center}
\caption{\label{fig:jpmfig} (Colour online) {\bf Left:} Currents in the two channels for a system $\alpha=0.75$, $\beta_+=0.25$, $\beta_-=0.75$, and with varying $p$. Solid lines show the current in the positive branch obtained from the mean-field analysis, dashed lines the current in the negative branch. {\bf Right:} Total current at $\beta_-=0.75$ for different choices of $\alpha$ and $\beta_+$. Lines are from theory. Symbols in both panels are from simulation ($L=100$), run for $10^4$ sweeps of the system, averaged over $50$ independent runs.}
\end{figure}

 In order to test the quantitative predictions of our mean-field approach for the total current we present a comparison with numerical simulations of the microscopic process in this section. In Fig. \ref{fig:jpmfig} we show the outcome of the mean-field expressions for the total current and the two currents in the individual branches for several choices of the model parameters. Lines in the figure are obtained from the theory, markers indicate corresponding results from numerical simulations. The left panel of Fig. \ref{fig:jpmfig} shows a vertical cut through the phase diagram depicted in the lower right panel of Fig. \ref{fig:pg1}, at a fixed value of $\beta_-=0.75$ and varying $p$. As $p$ is increased from zero to one, the system is first in the LM phase at low values of $p$, then enters the LL phase, and finally the HL phase. As seen in Fig. \ref{fig:jpmfig} direct measurements of the currents $j_\pm$ in the two branches of the system confirm the validity of the mean-field predictions to a good accuracy. The right-hand panel of Fig. \ref{fig:jpmfig} shows results for the total current, again comparing mean-field predictions and actual measurements and testing several combinations of the model parameters. Red circles show the total current corresponding to the model parameters of the left-hand panel. Green squares are for a symmetrical arrangement ($\beta_\pm=0.75$), corresponding to a vertical cut in the upper right panel of Fig. \ref{fig:pg1}, and blue triangles for a highly congested system with high injection rate ($\alpha=0.99$), and one `fast' exit route ($\beta_-=0.75$), combined with a `slow' alternative branch with a rather low ejection rate ($\beta_+=0.1$). This corresponds to the phase diagram shown in the right-hand panel of Fig. \ref{fig:pg2}. In this last situation the control parameter $p$, controlling the decision of incoming particles to move into the right and left branches respectively, needs to be tuned rather carefully to obtain an optimal flux through the system.

\begin{figure}[t]
\vspace{0em}
\begin{center}
\includegraphics[width=.8\textwidth]{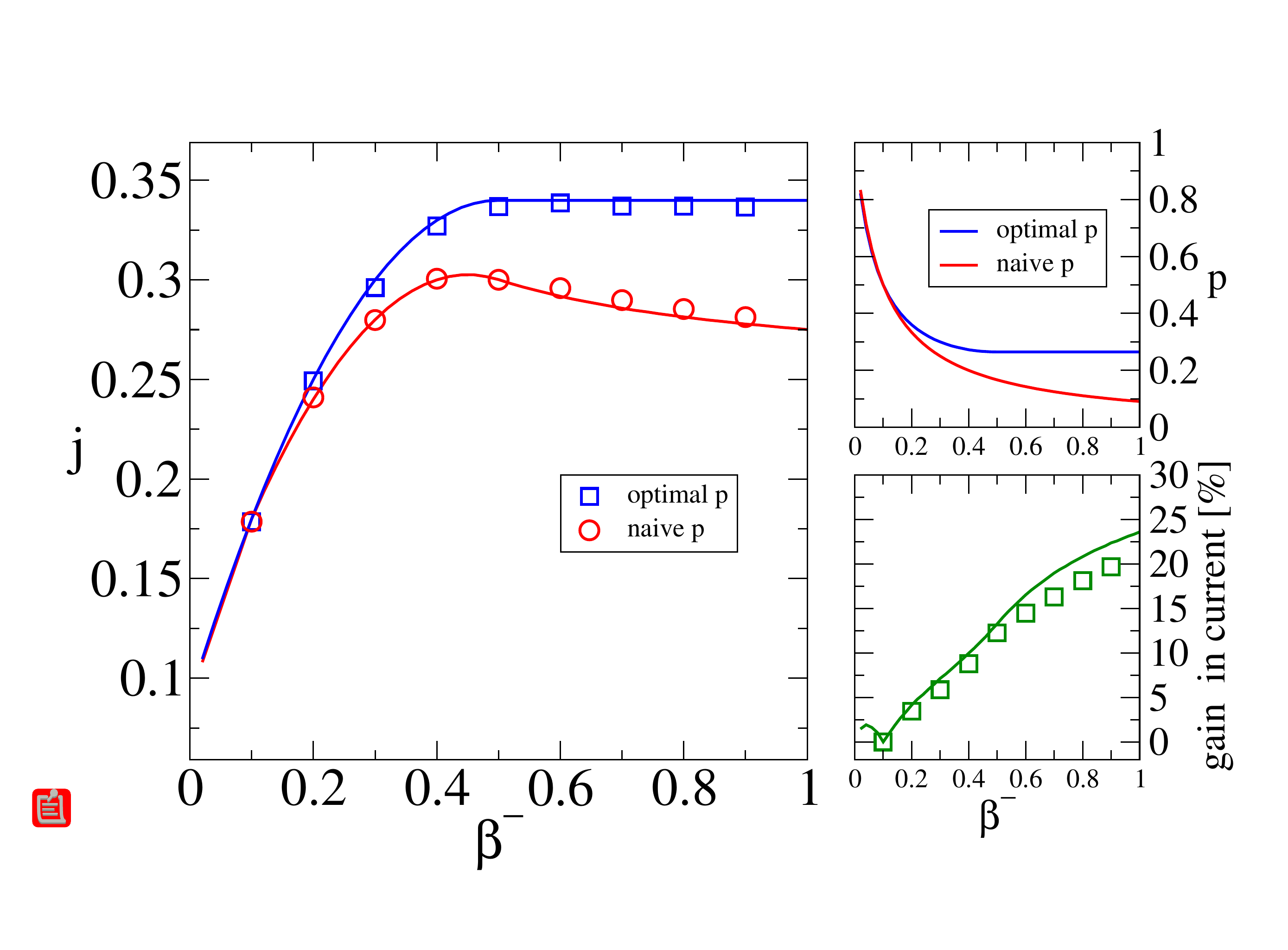}  
\end{center}
\caption{\label{fig:bestp} (Colour online) Naive versus optimised control policy for a system with $\alpha=0.9, \beta_+=0.1$. See text for further details of the definition of these policies. {\bf Main panel:} Total current for the naive and the optimized control policy. {\bf Upper right:} Naive and optimized values of $p$ as a function of $\beta_-$. {\bf Lower right:} Total gain in percent when the optimized strategy is used as opposed to the naive one (see text for detailed definition of the gain). Lines in all panels are from the mean-field theory, markers are from direct simulations of a system with $L=100$, run for $10^4$ sweeps, averaged over $50$ independent samples.}
\end{figure}

We investigate the effect of tuning the control policy in more detail in Fig. \ref{fig:bestp}. We have here again chosen a situation with high injection rate (albeit slightly less extreme that in the previous figure), and one very slow exit route, $\beta_-=0.1$, varying the ejection rate $\beta_+$ of the remaining branch. We then consider different control policies, parametrized by the choice of $p$. It is here important to stress that we restrict the analysis to {\em static} choices of the control policy, i.e. $p$ cannot be changed in time depending on the current state of the system. Instead we make an adiabatic assumption, and consider the stationary state of the system at fixed $p$. 

For each combination of model parameters $\alpha,\beta_\pm$ one can find the optimal (mean-field) control policy by choosing $p\in[0,1]$ such that the mean-field total current is maximised. We will refer to this as the `optimized' policy. A `naive' controller on the other hand may simply choose to direct incoming individuals into the two respective branches in proportion to the branches' ejection rates, again assuming that these are known. I.e. he or she may simply set $p=\beta_+/(\beta_++\beta_-)$. We will refer to this as the `naive' choice of $p$. 

Fig. \ref{fig:bestp} provides a detailed comparison of the naive and optimised control policies for the specific choice $\alpha=0.9, \beta_+=0.1$, and varying $\beta_-$. As seen in the figure the difference between the resulting fluxes is small for small values of $\beta_-$, i.e. when the two branches are very similar. This is also seen in the upper panel on the right-hand side of Fig. \ref{fig:bestp}, where we plot the values of $p$ for the naively obtained policy along with the optimal $p$. At small values of $\beta_-$ the difference between the two is rather small. As the ejection rate of the negative branch, $\beta_-$, is increased however, the situation becomes more and more asymmetric, and an optimised choice of $p$ can lead to a significant improvement over the outcome of the naive control policy (see main panel). The lower panel on the right highlights this situation, we show the total gain in percent when the optimised policy is used, as opposed to the naive one. More precisely, we plot $g=100\times (j_{\mbox{\small opt}}-j_{\mbox{\small naive}})/j_{\mbox{\small naive}}$. As seen in the figure optimizing the control policy can result in an overall improvement of up to $20\%$ or so, especially when the channels are very different in their ejection rates. We would like to stress though, that this appears only to be the case for congested systems with high injection rate. As a final remark we point out that the agreement between mean-field prediction and numerical simulation is rather good, even though not perfect. The mean-field approach tends to underestimate the current using the naive policy, and to overestimate it in the optimised case, leading to slightly inflated predictions for the overall gain. Still the error in prediction seems to be rather small (see lower panel on the right of Fig. \ref{fig:bestp}). We attribute the differences to a combination of finite-size effects and the assumptions underlying the mean-field approach. Small deviations of this type are also seen in the right-hand panel of Fig. \ref{fig:jpmfig}.

\section{Dynamic intervention}\label{sec:dyncontrol}

In this section we consider a dynamic intervention policy. I.e. the control parameter $p$ is now not constant in time, but instead its value can be modified in time depending on the current state of the system. Specifically we have considered a very simple control policy in which the value of $p$ depends only on the local densities of particles in each of the two branches near the central region. More precisely for a given depth $D$ we define
\be
\widetilde \rho_{D,+}=\frac{1}{D}\sum_{i=1}^Dn_i(t), ~~~ \widetilde \rho_{D,-}=\frac{1}{D}\sum_{i=-D}^{-1}n_i(t),
\ee
and then set 
\be
p(t)=\frac{1}{2}\left[1+\tanh[\Gamma(\widetilde \rho_{D,-}(t)-\widetilde \rho_{D,+}(t))\right],
\ee
where $\Gamma$ is a sensitivity parameter. For $\Gamma=0$ the value of $p(t)$ does not depend on the measured densities $\widetilde \rho_{D,\pm}$, for larger values of $\Gamma$ particles are increasingly sent into the less populated branch. For $\Gamma\gg 1$ we have
\be
p(t)=\left\{\begin{array}{ll} 1 & \widetilde \rho_{D,-}(t)>\widetilde \rho_{D,+}(t) \\ 0 &\widetilde \rho_{D,-}(t)<\widetilde \rho_{D,+}(t) \end{array}\right.,
\ee
i.e. particles strictly travel into the branch with the lower local density near the centre. The general setup is here reminiscent of the minority game \cite{challetzhang}, in which agents have the choice between two binary options, and where the minority group wins.

\begin{figure}[t]
\vspace{-1em}
\begin{center}
\includegraphics[width=.6\textwidth]{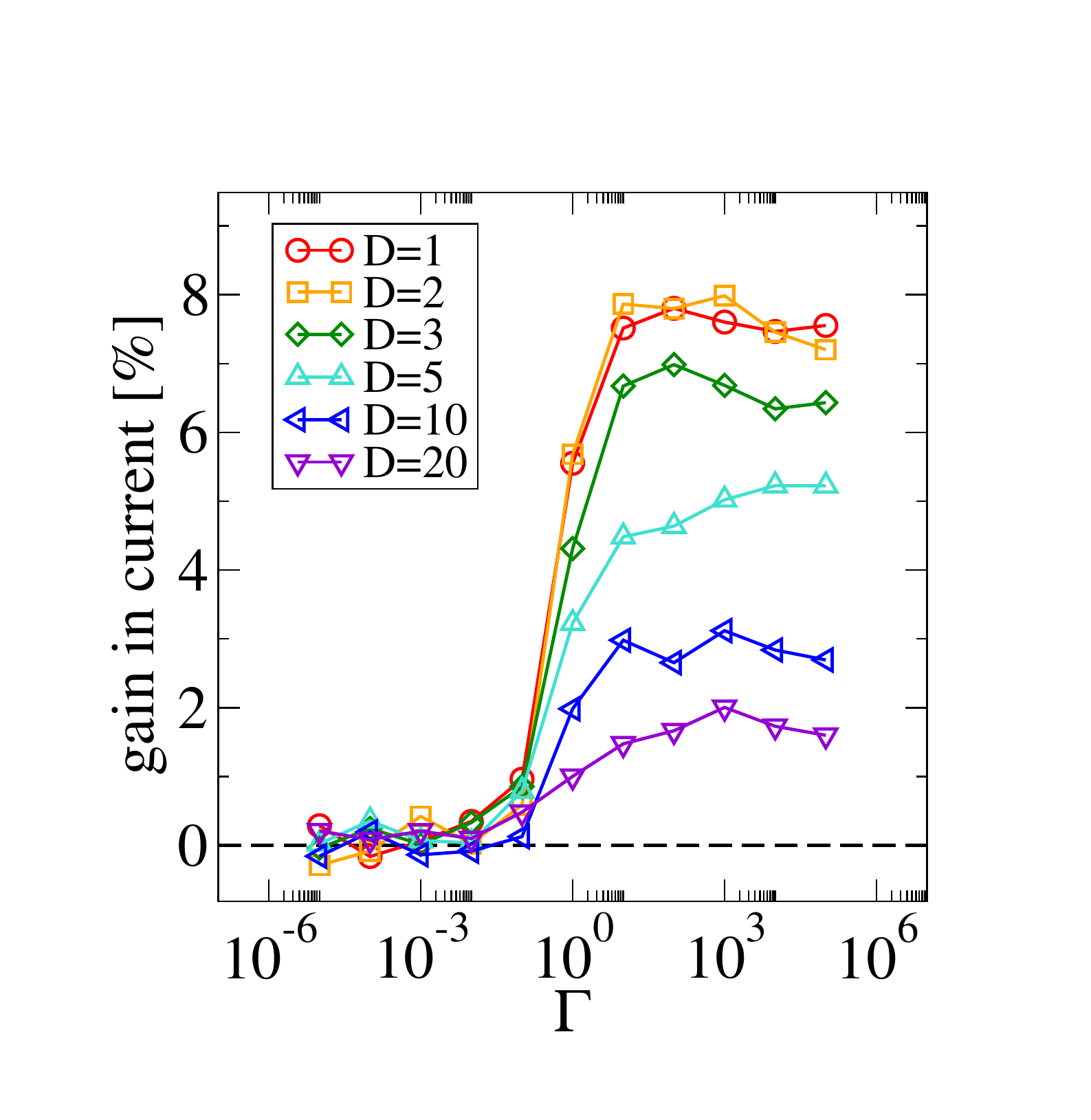}
\end{center}
\vspace{-1em}
\caption{\label{fig:dyn} (Colour online) Gain in total current for different dynamical intervention policies, $p(t)=0.5\left[1+\tanh[\Gamma(\widetilde \rho_{D,-}(t)-\widetilde \rho_{D,+}(t))\right]$, as explained in the text. Densities are measured within a depth of $D$ cells from the central cell. Remaining model parameters are $\alpha=0.8$, $\beta_1=\beta_2=0.3$. Data are from simulations with $L=100$, run for $10^4$ sweeps, averaged over $50$ runs.}
\end{figure}

Simulations are shown in Fig. \ref{fig:dyn}. The director variable $p(t)$ is here adapted continuously after each sweep. We choose a relatively high injection rate $\alpha=0.8$, and  a symmetrical setup with $\beta_+=\beta_-=0.3$ so that the optimal static intervention policy as predicted by the mean-field approach is given by $p=1/2$ by symmetry. As shown in Fig. \ref{fig:dyn} the total current through the system, i.e. the number of particles being removed from the end of the two branches per unit time, can be increased by about $8\%$ for our choice of model parameters, and if the control parameters $\Gamma$ and $D$ are chosen appropriately. The gain in percents is here defined by $100\times (j_d-j_{1/2})/j_{1/2}$, where $j_d$ is the total current obtained with dynamical intervention, and where $j_{1/2}$ is the current at constant $p=1/2$. As seen in the figure the sensitivity parameter $\Gamma$ has to exceed a certain threshold of about unity in order for dynamic control to have any effect. In addition to this the depth $D$ over which the densities in the two branches near the junction is measured needs to be chosen with care. If $D$ is too large, i.e. if densities are measured over the whole extent of the respective branches, then dynamic intervention has no significant effect. In fact, it is only the density of particles immediately ahead of the incoming particle that should be considered, we find optimal results for $D=1$ and $D=2$.
\section{Summary and outlook}\label{sec:summary}
In summary we have studied an abstract model of route choice in an evacuation scenario, based on a system of two totally asymmetric exclusion processes coupled through one common site. Using the known solutions of the TASEP dynamics the phase behaviour of the coupled system can be worked out on the mean-field level, and analytical predictions for the current through the system can be obtained. In most phases these results are in explicit closed form, in other phases the resulting equations need to be solved numerically. As a function of the model parameters the two-channel system exhibits complex phase behaviour with several topologies of phase diagrams observed. Testing the mean-field predictions of the total current against simulations generally gives good results, which is to some extent surprising, as the mean-field approach is based on severe assumptions and ignores correlations between different cells of the underlying automaton. When systematic quantitative deviations from the mean-field behaviour are found, then these are usually small, and they do not seems to restrict the ability of the analytical theory to identify optimised control policies. We have shown that optimising the control variable $p$, directing incoming individuals into either branch of the system, can result in a significant improvement of the evacuation speed, as measured by the total current running through both escape routes. These improvements are mostly observed for the case of congested systems with high injection rate, and strong asymmetry between the exit routes. We have also tested simple dynamical intervention policies, based on the time-varying densities of particles  in each of the two branches and measured in a region near the central entrance site. We show that this can lead to further improvement.

The TASEP model we use to propagate individuals in space is of course a rather stylised model, and we do not claim to have studied a `realistic' model of evacuation dynamics. However, as explained in the introduction, simple models can help to understand the basic mechanisms at work and to reveal representative types of behaviour. Cellular automata of vehicular traffic have for example been seen to qualitatively reproduce the fundamental diagram of road systems\cite{nagel, helbing}. Models of this type are also in use to model the movements of pedestrians. In addition to choosing more detailed `rules of engagement' on the microscopic level, possibly at the expense of analytical tractability, there is a number of other improvements one can make to the model studied here. For example one may give agents the opportunity to revise their initial choice of exit route, if they find that the one chosen initially is congested. These decisions may also rely on potential communication between agents. Work along these lines is currently in progress. A further route of extending the model might be to allow for more complex spatial arrangements beyond the simple case of two exit corridors. One may for example consider a system with multiple corridors, all coupled through one central exit, or other arrangements with a network of exit routes. In such cases the phase diagram of the model can be expected to be more complicated, with the total number of phases growing exponentially in the number of corridors in the system. In addition the optimisation problem becomes multi-variate as there is now one director variable per junction (or or multiple director variables for junctions with more than two exit routes).
\section*{Acknowledgements} This work is partially funded by an RCUK Fellowship (RCUK reference EP/E500048/1), the author acknowledges support by EPSRC (IDEAS Factory - Game theory and adaptive networks for smart evacuations, EP/I005765/1) and would like to thank Michalis Smyrnakis, Jamie King and Nick Jones for helpful discussions.

\end{document}